# The OBDT-theta board: time digitization for the theta view of Drift Tubes chambers.

*J. Sastre* [a]*, C.F. Bedoya* [a]*, S. Cuadrado* [a]*, J. Cuchillo* [a]*, D. Francia* [a]*, C. de Lara* [a]*, A. Navarro* [a]*, R. Paz* [a]*, I. Redondo* [a]*, D. Redondo* [a]*.*

[a] *Centro de Investigaciones Energéticas, Medioambientales y Tecnológicas.,*
 *Av. Complutense, 40, 28040 Madrid - Spain*
 *E-mail:* javier.sastre@ciemat.es

*On behalf of the CMS muon group.*

We present the design and performance of the On-Board electronics for the Drift Tubes (OBDT) for the superlayer theta along the direction parallel to the beam-line, the new board built to substitute part of the CMS DT Muon on-detector electronics. The OBDT-theta is responsible for the time digitization of the DT chamber signals for the theta view, allowing further tracking and triggering of the barrel muons. It is also in charge of part of the slow-control of the DT chamber inner electronics in the theta view. Prototypes of the OBDT-theta board are under validation in different laboratories in CERN, as well as in demonstrator chambers installed in the CMS experiment. This allows evaluation of the full functionality of the boards in real conditions, showing very satisfactory results.

# 1. Overview.

The on-detector electronics of the CMS Muon DT chambers [1] will need to be replaced for the High Luminosity operation of the Large Hadron Collider (HL-LHC) due to the increase of occupancy and trigger rates in the detector [2], which cannot be sustained by the present system. The OBDT-theta boards, together with OBDT-phi boards, will be located inside minicrate mechanics attached to the DT chambers at the CMS experiment. The OBDT-theta board will be in charge of performing the ~1 ns time digitization of the DT chamber signals of the theta view and the multiplexing for further transmission to the readout and trigger backend electronics. This board is also in charge of some of the slow-control tasks needed by part of the DT chamber system. There will be 180 new boards representing roughly 1/5 of all the OBDT boards in the system.

# 2. OBDT-theta board description.

The OBDT-theta board is built around a Microchip Polarfire FPGA, responsible for the time digitization of up to 228 input signals. For that, it implements a deserialization method which runs at 640 MHz obtaining a time bin of 0.781 ns. The input data is forwarded to the output optical link for data transmission to the readout and trigger chains, schematics shown in Fig. 1. Communication for this prototype is based on two VTRx+ transceivers [3] which provide two bi-directional links for slow-control and six transmitter links, capable of outputting data up to 10.24 Gbps to the backend system. One of the bi-directional links goes to the lpGBT chip [4] in the OBDT board, which provides the main slow-control functions and reception of the LHC clock and some TTC signals. The other one is directly connected to the FPGA, serving as a secondary slow-control, to recover OBDT-theta in case of loss of the main slow-control. The protocol implemented so far follows the lpGBT protocol for all links. The lpGBT chip plus a SCA chip in the board allow clock and synchronization reception, as well as e-link implementation for configuration and monitoring of the Polarfire FPGA. Through this link, it is also possible to implement the different slow-control functionalities of the barrel system, such as setting the front-end discriminators thresholds and bias values, implementation of the $I^2C$ links for temperature monitoring and channel masking, communication to the PADC (pressure analog to digital converters), RPC (Resistive Plate Chambers) and CMS muon Alignment slow-control chains and finally, control of the test pulse generation mechanisms that allows to perform the DT chamber time measurements calibration for the theta view superlayers.

Figure 2 shows the different firmware blocks of the Polarfire FPGA. A reference clock is received from a dedicated port of the lpGBT, which also sends a faster clock (320 MHz) and slow control information to the FPGA through the e-link port. Input hits are received from the inner chamber electronics and are digitized at the TDC block, using the same design as in the OBDT prototype [5]. On the other hand, the FPGA outputs Testpulse signal to the inner chamber electronics and the output data that is sent formatted in a lpGBT protocol to the next level of the electronics chain to the backend. It is to be noted that the synchronization commands from the Timing and Control Distribution System (TCDS) system, such as Bunch Crossing 0 (BC0) and Orbit reset (OC0), are received from the e-link protocol and decoded in the Timing and Trigger Control (TTC) block.

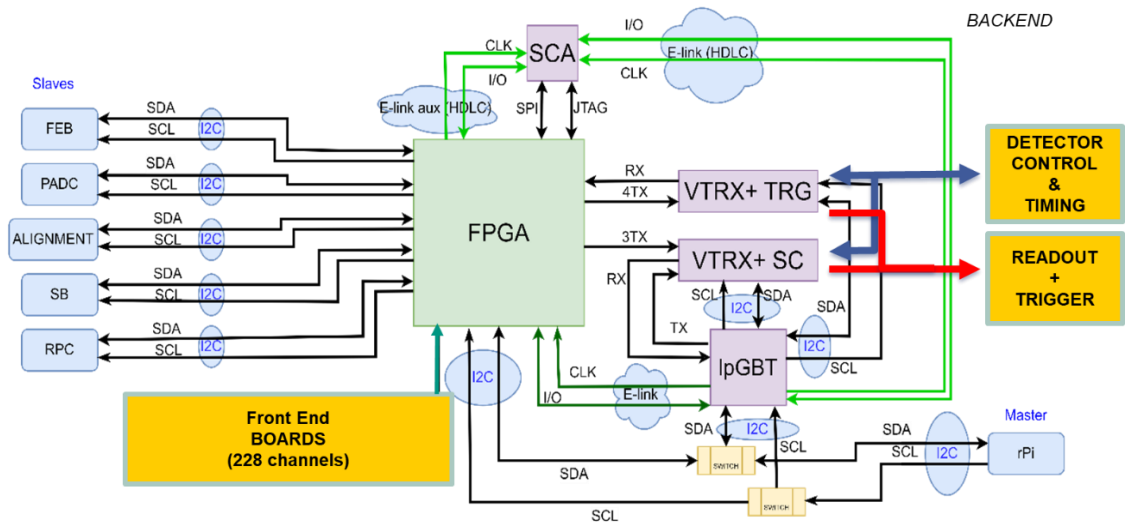

**Figure 1.** Diagram of the OBDT-theta main functional blocks.

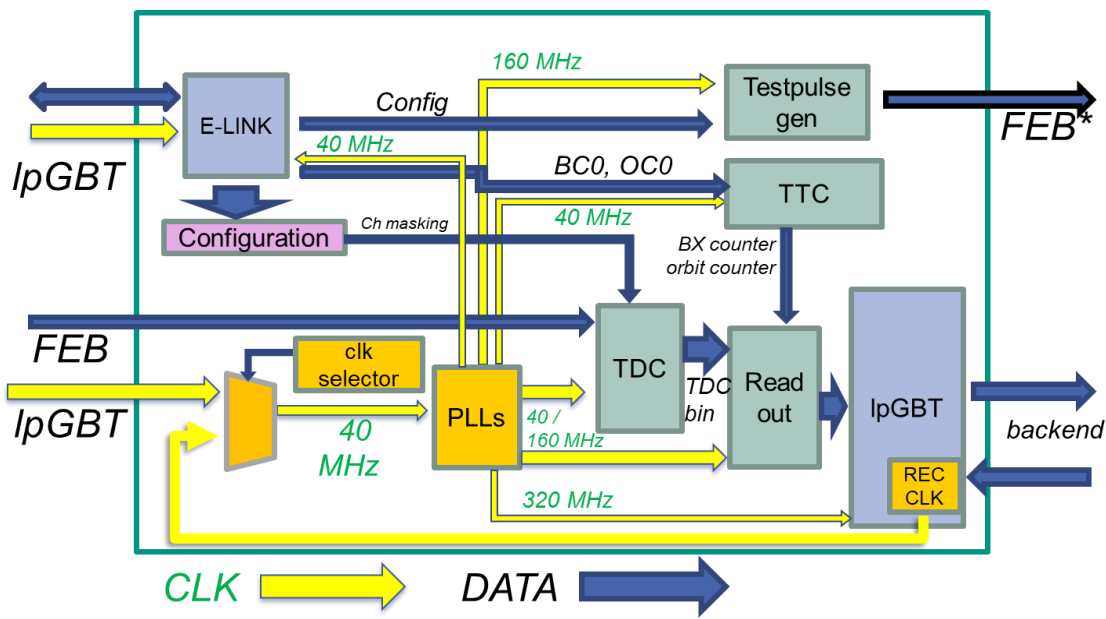

**Figure 2.** Diagram of the PolarFire FPGA firmware blocks of the OBDT-theta board.

In addition, a secondary link from the detector control & timing backend is connected directly to the FPGA transceivers, where the LHC clock is recovered and sent to a clock multiplexer driven by a clock selection logic, so it could be used in case the primary clock from lpGBT is lost.

## 3. Validation of the OBDT-theta.

Several validation tests have been performed with OBDT-theta prototypes. The first ones correspond to the measurements of the timing performance and, in particular, the Differential Non Linearity (DNL) of the different input channels, which were performed by injecting a non-correlated random signal. The results, depicted in Fig. 3, show that the DNLs are within the requirements of +-15%. DNL sawtooth pattern is believed to be caused by degradation of the DDR clock duty cycle along its routing path.

Figure 4 shows the performed measurements of channels crosstalk, where the DNL of one channel was measured with and without a neighboring channel having a signal injected. As can be seen, the effect of the neighboring signal is visible but it is below ±0.5 %, mainly due to higher proximity between certain pairs of differential signal in the PCB.

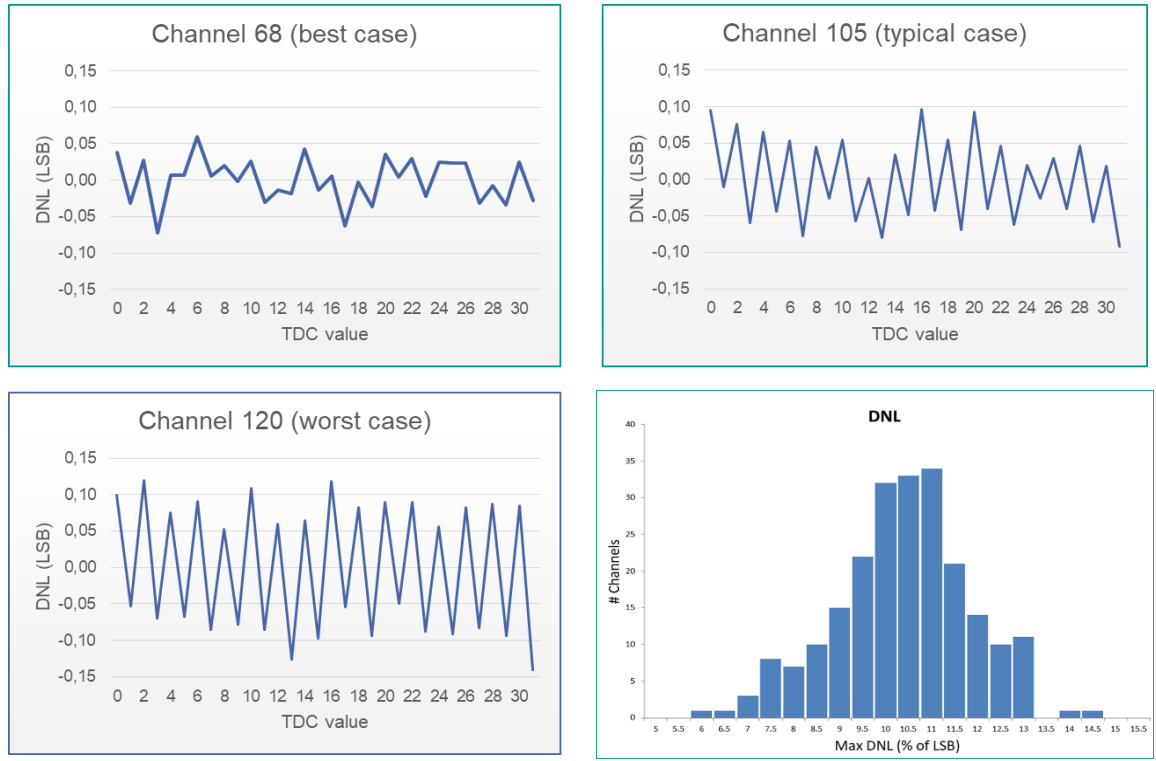

**Figure 3.** Images of some DNL measurements performed in different channels of the OBDT-theta board, and DNL channel distribution (down right).

Regarding time stability, the effect of power cycling has been analyzed, as shown in Fig. 5. For every channel, the difference between average time before and after power cycling at fixed intervals of 120 seconds is calculated, obtaining a maximum time drift of 50 ps. This effect can be only observed in channels whose digitization is done near the edge of the bin. Nevertheless this effect is not accumulative between power cycles.

Concerning signal integrity, some eye diagrams have been plotted using different hardware as shown in Fig. 6. In addition, a special Bit Error Ratio test have been performed for every link, obtaining excellent results: BER < $2{,}71 \cdot 10^{-14}$ / Inefficiency < $6{,}94 \cdot 10^{-13}$ (254 bits/word) for a test duration time of 10 hours ($1{,}44 \cdot 10^{12}$ packets sent).

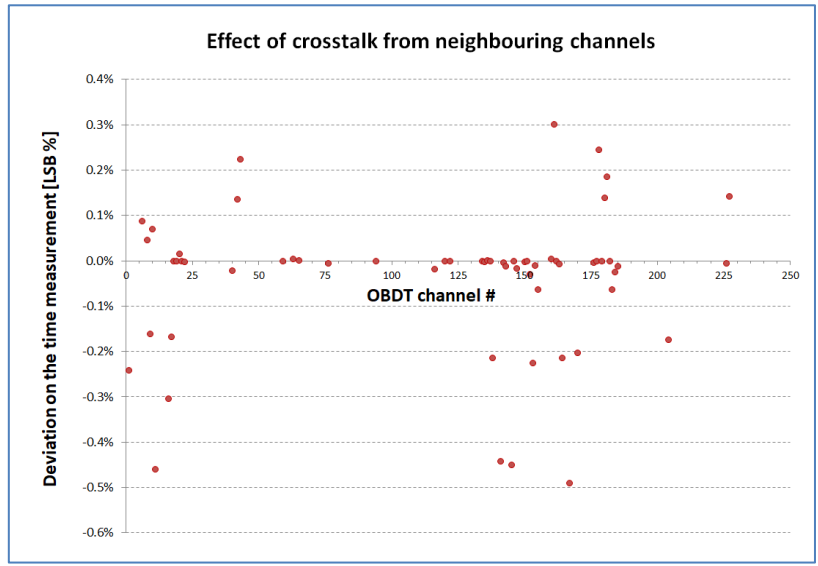

**Figure 4.** Effect of crosstalk in TDC time digitization.

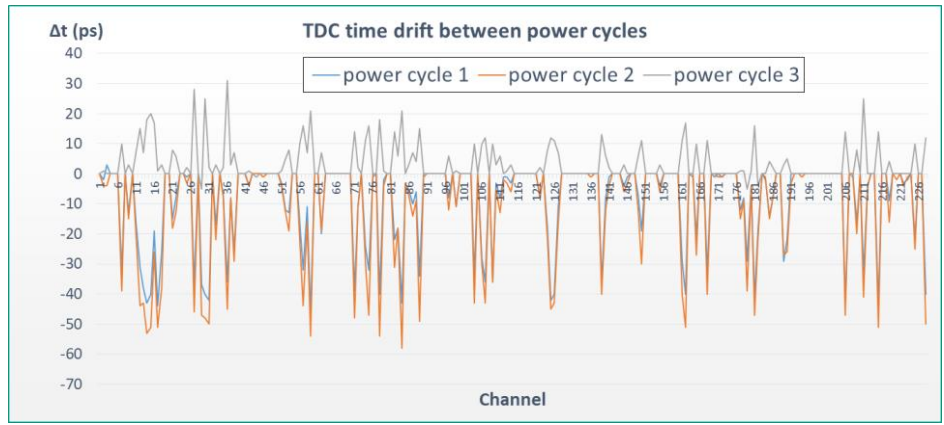

**Figure 5.** Effect of power cycling the OBDT-theta in time stability.

Irradiation tests of other versions of OBDT (OBDTv1 prototype at CERN CHARM [6], OBDTv2 phi in medical accelerator) suggest that Microchip Polarfire can be used with the intended functionality for the required dose below 2 Gy [2].

The present OBDT-theta prototype has been recently irradiated up to 50 times HL-LHC. The board did not require periodic power cycles or configurations to operate correctly, as it was the case with commercial optics ported by OBDTv1. No intervention was required until reaching a dose equivalent to 15 x HL-LHC for our application. The optocoupler ACPL-227, part of the power and safety system, was degraded during irradiation until it was not possible to operate the board > 30 Gy, which is more than 15 times the maximum expected dose.

Finally, other critical tests, related to the selection of the thermal pads that will keep in contact different devices (VTRx+, FPGA…) with the cooling frame where the board will be installed, suggest the use of SARCON-200GR-HD material [8].

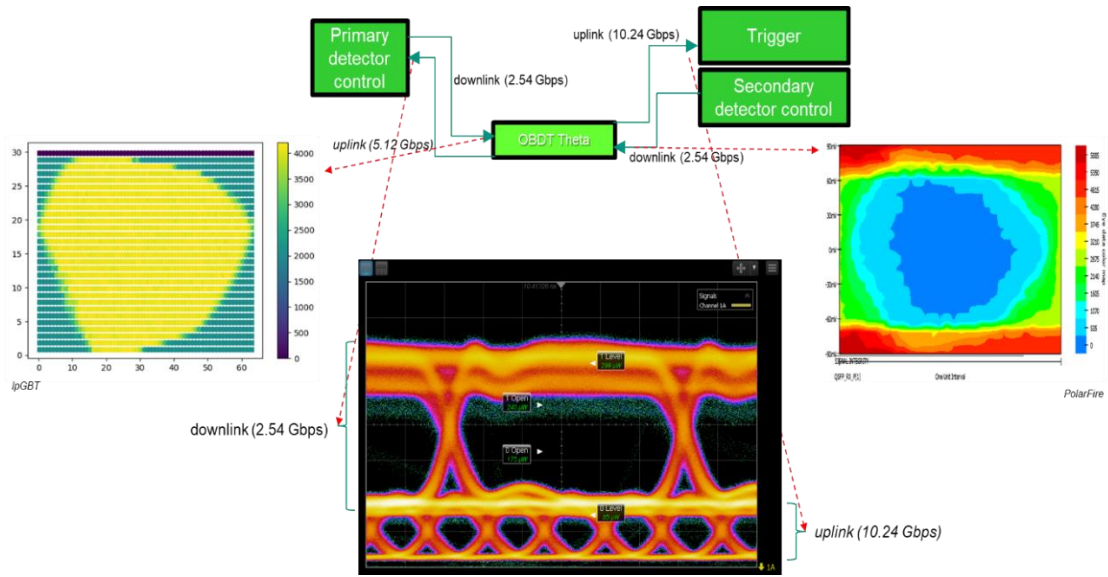

**Figure 6.** Eye diagrams of different links at the system.

## 4. Summary

Currently, 2 OBDT-theta boards are installed in chambers at CERN, which are used for the Slice Test in Point 5 (CMS experiment), completing a full chain of cosmic and collision data taking as well as calibration test pulse runs. The OBDT-theta prototypes show very good performance, with satisfactory validation tests results.